\documentstyle[epsf,12pt]{article}
\def\beq{\begin{equation}}
\def\beqn{\begin{equation*}}
\def\brr{\begin{array}}
\def\err{\end{array}}
\def\eeq{\end{equation}}
\def\eeqn{\end{equation*}}
\def\bea{\begin{eqnarray}}
\def\eea{\end{eqnarray}}
\def\bcenter{\begin{center}}
\def\ecenter{\end{center}}
\def\bright{\begin{flushright}}
\def\eright{\end{flushright}}
\def\bleft{\begin{flushleft}}
\def\eleft{\end{flushleft}}

      \def\Tr{{\rm Tr}}
      
\def\Det{{\rm Det}}       
\def\ni{\noindent}
\def\wt{\widetilde}      
      \def\ol{\overline}
\def\nn{\nonumber}

            \def\im{\mbox{Im}\, }

\def\braket#1#2{\,\langle\, #1 \,\vert\, #2 \,\rangle}
\def\com#1#2{\left[\, #1 \,,\, #2 \,\right]}
\def\acom#1#2{\left\{\, #1 \,,\, #2 \,\right\}}

\def\killing#1#2{\left(\, #1 \,,\, #2 \,\right)}

\def\to{\rightarrow}      \def\too{\longrightarrow}
     
\def\G{{\cal G}}            \def\E{{\cal E}}      
\def\A{{\cal A}}                  
           \def\mani{{\cal M}}      
      \def\calo{{\cal O}}            
\def\N{{\cal N}}

\def\Z{{\bf Z}}
\def\R{{\bf R}}
\def\C{{\bf C}}

\def\CS{{\scriptstyle {\rm CS}}}

\def\bz{{\bar z}}      \def\bw{{\bar w}}

\def\NP{ Nucl. Phys.}            \def\PR{ Phys. Rev.}            
      \def\PL{ Phys. Lett. }
   \def\CMP{ Commun. Math. Phys.}
      \def\IJMP{ Int. J. Mod. Phys.}

\def\ADM{ Adv. in Math.}         \def\ATMP{ Adv. Theor. Math. Phys.}

\def\JSM{ J. Sov. Math.}

\def\JHEP{ J.High Energy Phys.}

\begin{document}
\baselineskip=20pt

\begin{titlepage}
 \thispagestyle{empty}

\bleft {\small 
KUCP--0120\\[-5pt]
hep-th/9808094
} \eleft
 \vspace{60pt}

\bcenter
{\Large{\bf Classical Hamiltonian Reduction \\On $D(2|1;\alpha)$ Chern-Simons 
Gauge Theory \\and Large $N=4$ Superconformal Symmetry
}}
\ecenter

 \vspace{0pt}

\bcenter
{{\large Yukitaka Ishimoto}}\footnote{e-mail:
  ishimoto@phys.h.kyoto-u.ac.jp, fax: +81 (0)75-753-6804} 

{\it
 Graduate~School~of~Human~and~Environmental~Studies \\
 Kyoto~University,~Yoshida-Nihonmatsu-Cho,~Sakyo-Ku,\\
 ~Kyoto~606-8501,~JAPAN
}
\ecenter

   \begin{abstract}
3d Chern-Simons gauge theory has a
strong connection with 2d CFT and
link invariants in knot theory.
We impose some constraints on the $D(2|1;\alpha)$ CS
theory in the similar context of the hamiltonian reduction 
of 2d superconformal algebras.
There Hilbert states in $D(2|1;\alpha)$ CS theory
are partly identified with characters of the large $N=4$ SCFT 
by their transformation properties. 
   \end{abstract}

 \vskip3cm

\end{titlepage}

\section{Introduction}
\label{sec:00}

A decade before, a great deal of insights on 2d-3d field theoretic 
correspondence was exposed by Witten\cite{witten}.
In his celebrated paper, it is indicated that there is a certain
relation between 2-dimensional rational 
conformal field theory (2d RCFT) and
3-dimensional Chern-Simons gauge theory (3d CSGT).
This correspondence, called Chern-Simons-Witten theory (CSW), makes it 
possible to create not only topological invariants as a set of Jones
polynomials, also new link invariants of knot theory\cite{knot}. 
This means effectiveness of field theoretic methods on knot theory 
and has promoted a large amount of studies on this
subject\cite{MS,BN,ramallo,osp,link}.

Most of these works have not dealt with 
2d superconformal field theories (SCFT's) and applications on them, 
which recently developed by Ennes et al.\cite{osp}.
Here we formulate the $D(2|1;\alpha)$ CSGT and present a way to  
its connection with
2d large $N=4$ SCFT, adopting the idea of hamiltonian
reduction technique (HR method)\cite{BOoguri,ito}, which is well-known in 
2d CFT.

On the other hand, 1+1-dimensional $N=4$ SCFT's were shown to be 
good tools for describing a low energy structure of IIB string theory
compactified on $K3$ to six spacetime dimensions.
It has also been conjectured by Witten et al. and confirmed in the system
of $D1+D5$ brane configurations\cite{WS}.
In addition, 
among recent studies of $AdS_{d+1}$/$CFT_d$ correspondence on string theory,
there is such a 2d-3d correspondence on $AdS_{3}$ string theory
background as CSW theory reveals\cite{maldacena}.
It seems to follow that superalgebra-valued CSGT's are associated with 
operators of the boundary SCFT through the hamiltonian reduction method.
Though $AdS_{d+1}$/$CFT_d$ correspondence is not directly connected to
our case, these should nevertheless be a part of our motivation.

In this paper we perform an anti-holomorphic quantization procedure on CSGT. 
It has been applied to the
$osp(1|2)$ case where it leads to the $N=1$ SCFT\cite{osp}, 
but not to the cases of $N>1$ SCFT yet. 
In what follows,
we are going to extend this formalism to the platform 
with the basic Lie superalgebras $D(2|1;\alpha)$, which contains eight
fermionic generators, and then apply HR method to it.
Finally we explicitly show the partly correspondence between
2d large $N=4$ SCFT and 3d Lie superalgebra-valued Chern-Simons gauge 
theory(CSGT) in the context of HR method. It should be the basic
background for futher investigations on them, 
and there should be possibilities for future applications which is
shortly discussed in the end of this paper.
This formulation can also be applied to the other Lie
superalgebras\cite{ishimoto}.

\section{Anti-holomorphic quantization of CSGT}
\label{sec:1}

With an invariant bilinear form, 
$\killing{}{}$\footnote{The supertrace operation, seen in the basic
  definitions of superalgebras amounts to zero-killing form in this
  case\cite{kac}.},
defined in sect.\ref{sec:2}, we can write the action of
$D(2|1;\alpha)$ CSGT as 
\beq
\label{eq:CS}
   S_{\CS,k} = \frac{k}{4 \pi} \int_\mani \killing{A}{d A 
               + \frac13 \com{A}{A}} ,
\eeq
where $A$ is a $D(2|1;\alpha)$-valued one form over
an arbitrary three-dimensional $\mani$. $\com{}{}$ is a $\Z_2$-graded
commutator defined on the
superalgebra $D(2|1;\alpha)$. The variation of this action
leads to the Gauss law constraint
$F \approx 0 \,\,( F = d A + A^2 )$.
If $\mani$ has a boundary,
some 2-dimensional CFT is realized on it, but this is not our 
case.

There are two ways of formulations\footnote{Labastida et al. construct
  operator formalism in the context of the latter one\cite{ramallo}.} 
through quantization 
procedures on 2d Riemann surface $\Sigma$, which is made in our case by cutting
$\mani$ into two pieces.
One way is that: first the constraints due to the gauge invariance are
imposed, then quantization of the reduced phase space produces a
projectively flat vector bundle over the moduli space of complex
structures of $\Sigma$\cite{MS}. 
The other is called anti-holomorphic quantization procedure.
One constructs quantum mechanical wave functionals, then impose 
gauge constraints on the unconstrained wave functionals\cite{BN,ramallo}.
As we know, CSW theory arises in the
identification of conformal blocks of 2d
RCFT defined on $\Sigma$, with Hilbert states of 3d CSGT on $\mani$.
It is better to take the latter quantization scheme for our purpose,
since it enables us concretely to suppose that CSGT Hilbert states should be
SCFT characters, generating functionals for 2d current correlator blocks.

$$
\hbox{\raise 30pt \hbox{\epsfxsize=200pt \epsfbox{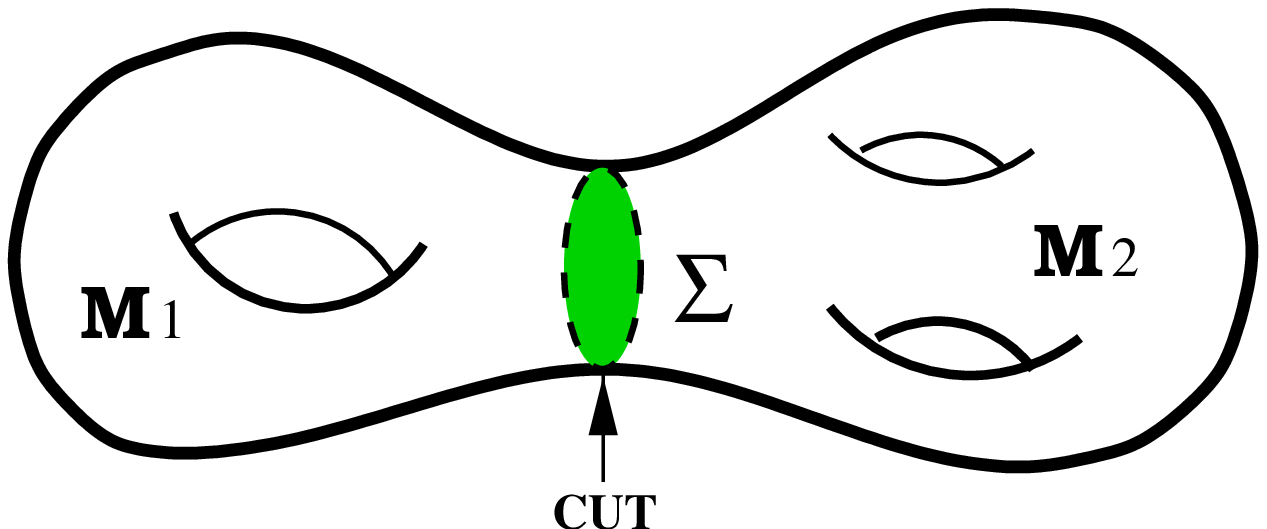}}}
\vspace{-45pt}
$$ 
\bcenter 
\label{fig:mani}{\small 
Fig.1: $\mani$ is divided into two
pieces which share a common \\2-dimensional Riemann surface $\Sigma$. }
\ecenter
\medskip

Let $\mani$ to be a manifold without boundary and cut it by
2-dimensional Riemann surface $\Sigma$ of genus $g$, then we get two
3-dimensional manifolds, $\mani_1$,$\mani_2$ ({\it Fig.1}). 
Each three-manifold has a boundary corresponding to the cut and  
each one can be identified with each other via homeomorphisms.
Taking $\mani$ locally to be $\R\times\Sigma$, we set $\R$ to be a time-axis
and impose the time-axial gauge $A_{t} = 0$.
In this set-up, we can canonically quantize the CSGT, introducing an
appropriate complex structure on $\Sigma$. It complexifies the gauge
group and sets 
$\com{A^a_z(z,\bz)}{A^b_\bw(w,\bw)} = {\pi}/{k}\,\,\delta^2(z-w)\,
{g^{ab}}/{2}$, 
where $\com{}{}$ is also $\Z_2$-graded one, $g^{ab}$ is a metric on
group manifold constructed from the bilinear form.
This canonical commutation relation induces 
a condition $A_z = A_\bz^\dag$.
We follow the anti-holomorphic quantization procedure developed by
Labastida\cite{ramallo}.
Let us assume that the states of Hilbert space $\Phi(A_\bz)$,
$\ol{\Phi}(A_z)$
on $\Sigma$ are spanned by
holomorphic and anti-holomorphic functions in terms of $A_\bz$,
$A_z \, (\equiv \ol{A_\bz})$.
Then anti-holomorphic quantization is accomplished by introducing an 
inner product on the functional space,
\bea
 \label{def:innpro}
  \braket{\Phi_2(A_\bz)}{\Phi_1(A_\bz)} = 
  \int DA_z DA_\bz \,\,e^{\frac{2\,k}{\pi}\int_\Sigma \killing{A_z}{A_\bz}}
  \,\,{\ol{\Phi_2(A_\bz)}}{\Phi_1(A_\bz)}.
\eea

Wilson line operators provide natural framework of
this topological quantum field theory, as gauge invariant observables.
These operators are gathered into the usual Feynman path integral
expression and have vacuum expectation value.
In the anti-holomorphic quantization we represent
these vev's as inner products on the unconstrained Hilbert
space, and impose the Gauss constraint on the states and gauge
invariance on the inner product as physical conditions.
One may realize that 
the corresponding conformal blocks of the SCFT satisfy these conditions.

\section{Conventions of the Basic Lie Superalgebra $D(2|1;\alpha)$}
\label{sec:2}

We are now in a position to fix the conventions of gauge field $A$ as an 
element of the superalgebra(SA) $D(2|1;\alpha)$.
Elements parameterizing the gauge field are defined by the use 
of root and weight systems, and Cartan matrix of the algebra.
The basic Lie superalgebra $\G \equiv D(2|1;\alpha)$ 
is $rank~(\G) = 3$ and dual Coxeter number $h^\vee = 0$. 
By definition, $\G$ is decomposed into two parts up to ${\bf Z}_2$-grading, 
$\G=\G_0 \oplus \G_1$.
The even subalgebra $\G_{0}$ is a Lie algebra
$\G_{0} = \A_1 \oplus \A_1 \oplus \A_1$, while the odd subalgebra $\G_{1}$
is set to be a fundamental representation of $\G_{0}$, 
in other words, $\G_{0}$ on $\G_{1}$ is $sl_2 \otimes sl_2 \otimes sl_2$.

In general, the basic Lie SA's are equipped with an invariant
bilinear form 
which induces an inner product on the root space. 
Even roots have positive or negative length squared, while 
odd ones have zero length squared.
It should be noted that, in contrast to the usual case of Lie algebras, 
Cartan matrix of Lie SA is not uniquely determined.
There are several possible inequivalent choices of simple 
roots. [ {For more information on Lie superalgebras see
  e.g.\cite{kac}} ]

In our case, simple roots are chosen to be
$\alpha_{(1)},\alpha_{(2)},\alpha_{(3)}$, and 
their inner products are given by 
\bea
\label{eq:simpleroots}
   \alpha_{(1)}^2 = 0 ,\, \alpha_{(2)}^2 = -2 \gamma ,\, \alpha_{(3)}^2 =
   -2(1-\gamma),
\nn\\
   \alpha_{(1)} \cdot \alpha_{(2)} = \gamma ,\, \alpha_{(1)}\cdot\alpha_{(3)} =
   1-\gamma ,\, \alpha_{(2)}\cdot\alpha_{(3)} = 0 ,
\eea
where $\gamma\equiv\frac{1}{1+\alpha}$ ($\gamma\not=0,\pm\infty$).
The positive even roots, conventionally denoted by $\Delta_0^+$, are
mutually orthogonal to each other,
$\alpha_+ \equiv \alpha_{(2)}$, 
$\alpha_- \equiv \alpha_{(3)}$, 
$\alpha_\theta \equiv 2 \alpha_{(1)} + \alpha_{(2)} + \alpha_{(3)}$, and 
the positive odd roots $\Delta_1^+$ are spanned by four elements,
$\beta_- \equiv \alpha_{(1)}$, 
$\beta_{-K} \equiv \alpha_{(1)} + \alpha_{(3)}$,
$\beta_+ \equiv \alpha_\theta - \beta_-$, 
$\beta_{+K} \equiv \alpha_\theta - \beta_{-K}$.
With these roots, we can set the even part of this algebra to be a real form
$su(2)_+ \oplus su(2)_- \oplus su(2)_\theta$. One of these three $su(2)$, 
explicitly denoted by $su(2)_\theta$, will be 
constrained later in the context of hamiltonian reduction.

A set of canonical basis is given by $\{ E_\alpha, e_\beta, h^i \}$
where $\G_0$ and $\G_1$ are generated by $\{ E_\alpha \,,\, h^i \}$ and 
$\{ e_\beta \}$, respectively.
Their commutation relations are given by 
\bea
\label{def:WCbasis}
   \com{E_{\alpha_i}}{E_{\alpha_j}} &=& \frac{2\,\alpha_i \cdot h }{\alpha_i^2}
   \delta_{\alpha_i + \alpha_j , 0}, \nn\\
   \acom{e_{\beta_\mu}}{e_{\beta_\nu}} &=& N_{\beta_\mu,\beta_\nu} 
   E_{\beta_\mu+\beta_\nu} + \beta_\mu \cdot
   h\,\,\delta_{\beta_\mu+\beta_\nu,0} 
   \ \ for \ \ \beta_\mu\in\Delta_1^+, \ \beta_\mu+\beta_\nu\in\Delta_0  ,\nn\\
   \com{e_{\beta_\mu}}{E_{\alpha_i}} &=& N_{{\beta_\mu}, {\alpha_i}} e_{{\beta_\mu} + {\alpha_i}} 
   \ \ for \ \ {\beta_\mu}+{\alpha_i} \in \Delta_1  , \nn\\
   \com{h^i}{E_{\alpha_j}} &=& {\alpha_j}^i\, E_{\alpha_j} , \quad
   \com{h^i}{e_{\beta_\mu}} \,=\, {\beta_\mu}^i\, e_{\beta_\mu} ,
\eea
where 
$\alpha_i, \alpha_j \in \Delta_0$, and 
$\beta_\mu, \beta_\nu \in \Delta_1$. 
The suffix $i$ of $\{h^i\}$ takes a value of $\{+,-,\theta\}$.
${\alpha_j}^i$ and ${\beta_\mu}^i$ are components in the $h^i$-direction.

On $D(2|1;\alpha)$, Killing form, naturally defined by supertrace, is a 
cumbersome degenerate, zero-Killing form.
Instead of the above definition one may replace it to a 
non-degenerate Killing form with a non-degenerate invariant bilinear form 
on $D(2|1;\alpha)$, letting the algebra contragredient.
Then the Killing form is defined by
\bea
\label{def:killing}
  \killing{E_{\alpha_i}}{E_{\alpha_j}} &=& \frac{2}{\alpha^2}
  \,\delta_{\alpha_i,-\alpha_j} ,\quad
  \killing{h^i}{h^j} \,=\, \delta^{ij} , \nn\\
  \killing{e_{\beta_\mu}}{e_{-\beta_\nu}}   
  &=& - \killing{e_{-\beta_\mu}}{e_{\beta_\nu}} 
  \,=\, \delta_{\beta_\mu,\beta_\nu}  \ \ \ \ for \ \
  \beta_\mu,\beta_\nu\in\Delta_1^+ .
\eea
This Killing form $\killing{}{}$ on $\G$ enables us to identify the dual
space $\G^{\vee}$ of $\G$ with $\G$ itself.

Now we can express an element of the algebra in terms of the canonical basis, 
\beq
\label{def:J}
   A(z,\bz) =  \sum_{\alpha\in\Delta_0} A^\alpha(z,\bz) E_\alpha
      + \sum_{\gamma\in\Delta_1} A^\gamma(z,\bz) e_\gamma
      + \sum_{i=+,-,\theta} A^i(z,\bz) \wt h_i , 
\eeq
where the Cartan subalgebra of $\G$ is spanned by $\{\wt h_i\}$
($\wt h_i \equiv \frac{\alpha_i \cdot h^i}{(\alpha_i)^2}$).
Infinitesimal gauge transformation of the $D(2|1;\alpha)$ current $J$ is 
written down as 
\bea
\label{def:Jgauge}
   \delta_{gauge} A(z,\bz)  &=& [ \Lambda(z,\bz) , A(z,\bz) ] + \partial \Lambda(z,\bz),\nn\\
\label{def:gauge}
   \Lambda(z,\bz) &=&   \sum_{\alpha\in\Delta_0} \epsilon^\alpha(z,\bz) E_\alpha
      + \sum_{\gamma\in\Delta_1} \epsilon^\gamma(z,\bz) e_\gamma
      + \sum_{i=+,-,\theta} \epsilon^i(z,\bz) \wt h_i . 
\eea
If we impose holomorphy on this current, central extension of
$\G$ is realized so that coadjoint action on its dual space provide the above gauge transformation, setting its level $k=1$ \cite{ito}.

\section{Determination of the Hilbert States in $D(2|1;\alpha)$ CSGT}
\label{sec:3}

In eq.(\ref{def:innpro}) we define the inner product on Hilbert space
of $D(2|1;\alpha)$ CSGT. The transition from this unconstrained
Hilbert space to the physical one is obtained by imposing the
spatial gauge invariance on $\Sigma$. This physical space satisfies 
the Gauss law constraint automatically. This is done in two steps. First 
a subspace of wave functionals is selected by the requirement that 
inner product be totally gauge invariant, which can generate some
constraints on the Hilbert states. 
Then this vector space is
endowed with the inner product by restricting the $A$-integration to a
subspace intersecting every gauge orbit once.

When $\mani_1(,\mani_2)$ is a solid ball ($i.e.$ $\Sigma=S^2$),
there is no new constraint on
the Hilbert space except for Gauss law constraint.
Thus, this case cannot contain so much information on the Hilbert space
that establishes connections with SCFT\cite{ramallo}.
In the following, we consider a genus-1 handlebody case as in
\cite{ramallo}, namely, $\mani_1(,\mani_2)$ is a solid torus and
$\Sigma=T^2$. 

Flat connection admits the following parameterizations of the gauge fields on $\Sigma=T^2$, 
\bea
 \label{eq:gauge}
  A_\bz = (u_a \,u )^{-1} \partial_\bz (u_a \,u) , \ \ 
  A_z = (u_a \,\bar{u} )^{-1} \partial_z (u_a \,\bar{u}) , \nn
\eea
where $u$ is a single-valued map: $\Sigma \to \G^\C$, and $u_a$
contains non-trivial global information associated with 
the fundamental group on $\Sigma$\cite{GK},
\bea
 \label{eq:ua}
   u_a (z,\bz) = \exp\left[\frac{{\rm i}\pi}{\im \tau}\left\{ \int^\bz 
   \ol{\omega(z^{\prime})} \,\, a\cdot {\wt h} 
   - \int^z \omega({z}^{\prime}) \,\,\bar{a}\cdot {\wt h} \right\} \right] .
\eea
The space of the one-forms over $\mani$ allows the Hodge parameterization 
where the (anti-)holomorphic one-form $\omega(z)$($\ol{\omega(z)}$) is 
taken to be $\int_\alpha \omega = 1, \ \int_\beta \omega = \tau,$
and 
$\int \omega\wedge\ol{\omega} = \im\tau$. $\alpha$ is a contractable
homology cycle in the solid torus and $\beta$ is a non-contractable one.
``$a$'' is a $3$-dimensional vector  
and ${\wt h}$ is in the Cartan subalgebra.
Contraction of them is introduced in such a way that 
${a\cdot{\wt h} = \sum_{i=\theta,+,-} a^i }{\wt h}_i$.
The condition $A^\dag_z = A_\bz$ causes relations on parameters,
$\bar{u}^{-1} = {u^\dag}$ and $u_a^{-1} = u_a^\dag$.

This parameterization induces a
change of the variables in Haar measure from $A_z, A_\bz$ into
$u_a$, $u_a^\dag$, $u$ and $\bar u$, denoted by $F$.
It generates the Jacobian $j( a, \bar a, u )$ as below.
\bea
\label{def:jacobian}
  F^* ( {\rm Haar~measure} ) &=& j ( a, \bar a, u) 
  d u_a d u_a^\dag du d\bar u ,\nn\\
  j ( a, \bar a, u)\!\!\!\!\!\!\! &=\!\!\!\!\!\!\!& C (\im \tau)^3 
   \frac{ \Det^{\prime} (\ol{D}_{A_\bz}^\dag \ol{D}_{A_\bz}) }
  {\det \killing{\epsilon_i}{\epsilon_j}\det\killing{a_i}{a_j}}, \nn
\eea
where 
$C$ arises from the normalization up to group manifold and
$\ol{D}_{A_\bz}$ is the map from $\G^\C$-valued function to
$\G^\C$-valued one-forms,
$$\ol{D}_{A_\bz}: \E\too\A \ \ , \ \ \ol{D}_{A_\bz}\epsilon =
\partial_\bz \epsilon + \com{A_\bz}{\epsilon}.$$
$\ol{D}_{A_\bz}^\dag$ is its adjoint map.
In order to explain the rest of the unknown variables,
some conventions should be introduced, 
$\epsilon\in\E$ ($\E$ is a space of $\G^\C$-valued functions on
$\Sigma$),
$a \in \A$ ($\A$ is a space of $\G^\C$-valued one-forms on $\Sigma$)
and kernels of the above two maps are denoted by $\ker \ol{D}_{A_\bz} =
\{\epsilon_i\}$, $ \ker \ol{D}^\dag_{A_\bz} = \{a_i\}$ ($i = +,-,\theta$). 
Matrix $\killing{\epsilon_i}{\epsilon_j}$ is defined by a
projection on the space $\ker \ol{D}_{A_\bz} = \{\epsilon_i\}$.
$\Det^{\prime}$ is $\zeta$-function regularized,
$\Det^{\prime} \calo = \lim_{\epsilon\to 0} \exp 
[ \int_\epsilon^\infty \frac{d t}{t} \,\Tr^{\prime} \exp\{ -t \calo\} ]$, 
where $\Tr^{\prime}$ is a trace over $\E$ excluding the kernels.
The above definition gives rise to the famous chiral anomaly\cite{GK,alvarez}
\bea
\label{eq:jacobian}
     \frac{ \Det^{\prime} (\ol{D}^\dag_{A_\bz} \ol{D}_{A_\bz}) }
     {\det\killing{\epsilon_i}{\epsilon_j}\det\killing{a_i}{a_j}} 
   &=& \frac{ \Det^{\prime} (\ol{D}^\dag_{A_\bz} \ol{D}_{A_\bz}) }
   {\det\killing{\epsilon_i}{\epsilon_j}\det\killing{a_i}{a_j}} \Bigg|_{u=const.}
   \!\!\!\!\!\! S_{G,2 h^\vee} ( u \bar{u}^{-1} , A_\bz |_{u=const.} ) \nn\\[5pt]
   &=& (\im \tau)^{-6} \Det^{\prime} (\ol{D}^\dag_{A_\bz} \ol{D}_{A_\bz}) 
   \big|_{u=const.} . 
\eea
In general, 
chiral anomalies are expressed by gauged WZW(Wess-Zumino-Witten)
action $S_{G,k}(g,A)$ with some gauge group $G$, its element $g$, 
a gauge field $A$, and its Kac-Moody level $k$. 
Note that in the second line of
eq.(\ref{eq:jacobian}), the gauged WZW action vanishes due to the dual
Coxeter number of $\G$ ($h^\vee = 0$). It means the absence of
the well-known quantum shift $k \to k+h^\vee$ in this theory.
It persists everywhere in what follows.

At this stage, we refer to the hamiltonian reduction and apply to our case.
The hamiltonian reduction on basic classical Lie SA's is 
realized as a constraint on lowering (or raising) current
of $A_n$ among (semi-)simple components of the
even subalgebras. As the constraint is retained along the whole gauge
orbit, gauge transformations lead to a Poisson bracket structure and
operator product expansion (OPE) relations of
superconformal algebra(SCA) or $W_n$ algebra.

Along this strategy, we introduce the analogous constraint for our case,
which is imposed on one of the $su(2)_\theta$ current 
in eq.(\ref{def:J}), so that $$J_{-\theta} = 1 .$$ 
This constraint on the gauge symmetry
turns out to be translated into constraints on the gauge parameters, 
$\Lambda$, by extracting the $J_{-\theta}$ part of the gauge
transformation.
\beq
\label{def:HR}
   \epsilon^{-\theta} = \epsilon^i \cdot \alpha_\theta = 0 \,,\, 
   \epsilon^{-\gamma} = 0 \ \ ( \gamma \in \Delta_1^+ ) .
\eeq
However,
with these constraints, there is still left the following gauge group $\N$ 
(equal to ${su(2)_+}\oplus{su(2)_-}\oplus{{\bf n}_1}$ ).
${\bf n}_1$ is generated by
\bea
   \Lambda_{{\bf n}_1} = \epsilon_\theta\, E_\theta +
   \sum_{\gamma\in\Delta_1^+} \epsilon_\gamma\, e_\gamma. \nn
\eea
Leaving the $su(2)_+ \oplus su(2)_-$ gauge symmetry, 
this residual gauge symmetry must be subtracted from the functional
integral eq.(\ref{def:innpro}). We will discuss this point later.
With an appropriate gauge fixing procedure for this residual
one, for example, Drinfeld-Sokolov gauge\cite{DS},
one can obtain the complete OPE relations isomorphic to the non-linear
large $N=4$ SCA in two dimensions\cite{GShwimmer}.

Now, it is clear how to impose HR constraints on our $D(2|1;\alpha)$ CSGT, 
on the inner product of its Hilbert states.
Let us realize eq.(\ref{def:HR}) as delta functions of HR constraints
and append them into 
the integrand of eq.(\ref{def:innpro}). Its restriction of the
gauge transformations determines properties of the Hilbert states.
Finite gauge transformation of the gauge field $A$ is represented by 
\bea
 \label{eq:gauge1}
  {}^g A = g^{-1} A g  + g^{-1} \partial g .
\eea
The HR method restricts the above $g$ to be an
exponential map with constrained parameters, 
\bea
 \label{eq:restriction}
    g &=& \exp[\Lambda{}] \,, \quad
 \label{eq:gaugeparam}
    \Lambda{} \,=\, \epsilon_\theta{} E_\theta +
    \sum_{\alpha\in\Delta_{\wt\G}} \epsilon_\alpha{} E_\alpha +
    \sum_{\gamma\in\Delta_1^+} \epsilon_\gamma{} e_\gamma +
    \sum_{i=+,-} \epsilon^i{} h^i .
\eea
where $\wt\G$ denotes $su(2)_+ \oplus su(2)_-$. Note that $\Lambda$
includes $\Lambda_{{\bf n}_1}$ corresponding to ${\bf n}_1$.
We apply this expression to all the gauge transformations.

The gauge transformation, eq.(\ref{eq:gauge1}), can be simply rewritten 
in terms of the $u$-parameter transformation, which we denote type (i) 
transformation as in \cite{ramallo},
$ u \too ug $. 
By this type (i) gauge transformation, the integrand apart from
a bilinear term of Hilbert states, is transformed and generates
additional factors so that
\bea
   e^{\frac{2\,k}{\pi} \int_\Sigma \killing{A_z}{ A_\bz}}
   &\too& e^{\frac{2\,k}{\pi} \int_\Sigma \killing{A_z}{A_\bz}} 
        e^{k \left(2 \Gamma(g) + \langle u_a u , g \rangle 
           + \ol{\langle u_a u , g \rangle} \right) } ,\nn
\eea
where $\Gamma(g)$ denotes WZW-action and 
$\langle u,g \rangle$ is given by
$ \langle u,g \,\rangle = \frac{2}{\pi} \int_\Sigma \killing{u^{-1}
  \partial_\bz u}{\partial_{z} g g^{-1}}$. 
Type (i) gauge invariance of the integrand leads to the next transformation properties of the Hilbert states,
\bea
   \Phi(A_{\bar z})
   &\too& e^{-k \left( \Gamma(g) + \langle u_a u , g \rangle
     \right)} \Phi(A_{\bar z}) , \nn
\eea
Thus, wave functional $\Phi(A_{\bz})$ is expected to be decomposed
into the following three parts
\bea
\label{eq:Phi}
  \Phi(A_{\bar z}) &=& \kappa \Psi_k (u_a u) \,\Xi(u_a) \nn\\
                   &=& \kappa \exp [ -k ( \Gamma(u) + \langle u_a,
                   u\rangle)] \Psi_k(u_a) \,\Xi(u_a) ,
\eea
where constant $\kappa$ and $\Xi(u_a)$ should be determined up to the
orthonormality of Hilbert states.

Substituting eq.(\ref{eq:jacobian}), eq.(\ref{eq:Phi}) into eq.(\ref{def:innpro}), 
HR constrained inner product is thus obtained, 
\bea
 \label{eq:innerprod.sol}
    \braket{\Phi_1}{\Phi_2}
    &=& \bar\kappa_1 \kappa_2 \int {du_a}{d\bar u_a} [K(\tau,a)]
    \exp\left[ -k \langle u_a , u_a^{-1} \rangle \right]
    \times \ol{\Psi_{1,k}(u_a)}\,
    \Psi_{2,k}(u_a) , \nn\\
 \label{eq:Konstant}
\qquad    [ K(\tau,a) ]\!\!\!\!\!\! &\equiv\!\!\!\!\!\!&\;
    (\im\tau)^{-3} \Det^{\prime}
    (\ol{D}^\dag_{A_\bz} \ol{D}_{A_\bz}) \big|_{u=const.}\,
    \ol{\Xi_1(u_a)} \,\Xi_2(u_a) \nn\\
    &&\quad\times 
    \int \frac{{du}{d\bar u}}{[{\rm gauge~volume~of~{{\bf n}_1}}]} 
    \exp\left[ -k \,\Gamma(u\bar u^{-1}, B) \right] 
    \delta({\rm HR\!\!-\!\!\,constraints}) , \nn\\
\qquad &&  \left( B_{\bz}\equiv {u_a}^{-1}\partial_\bz u_a \,,
           \quad B_{z} \equiv {u_a}^{-1}\partial_z u_a\right) .
\eea
In eq.(\ref{eq:Phi}), Hilbert states are decomposed into three parts,
$\Psi_k (u_a)$, $\Xi(u_a)$, and an exponential factor. This factor with
its anti-holomorphic partner and a $A_\bz$-$A_z$ potential term is
rewritten in the form of a gauged WZW action and a new potential term
in terms of $u_a$, $u_a^{-1}$. While the gauged WZW action is alone to be
integrated over $u$-variable, it has to be treated carefully.
The determinant at a point of a constant $u$ can be thought of a kind of
gauge fixing. Trivial $u=1$ gauge leaves only Cartan generators in
the form of $A_\bz$, $A_z$,
and easily enables us to construct an effective quantum mechanics,
integrating out the u-parameter dependence.
In this thought, $u$-parameter space is also regarded as a gauge orbit
going through the point of the constant $u$ and its integration 
must be restricted along the HR context of 2d SCFT by an appropriate
delta function. Assembling the results and discussions, $
[ K(\tau,a) ]$ could be the form in eq.(\ref{eq:Konstant}) and provide 
a part of large $N=4$ character. Moreover, it should be noted
that the gauge volume of ${\bf n}_1$ must be subtracted from
$u$-integration for getting a proper inner product, since integration
over ${\bf n}_1$ with delta function of HR simply gives rise to the 
gauge volume for ${\bf n}_1$-symmetry.

Next, 
in order to construct an explicit form of $\Psi_k(u_a)$, 
we consider another gauge transformation (ii)
\bea
 \label{eq:gauge2}
  u_a \too u_a \hat{g} ,\, u \too {\hat{g}}^{-1} u ,\nn
\eea
where $\hat{g}$ is a map $\hat{g}$:$\Sigma \to$[Maximal torus of the group],
parameterized by  
\bea
     \hat{g}_{n,m} = \exp \left[ \frac{ \rm{i}
     \pi}{\im\tau}\left\{ (n+m\tau)\cdot {\wt h} \int^{\bar z} \ol{\omega(z^{\prime})} 
     - (n+m\bar\tau)\cdot {\wt h} \int^z \omega(z^{\prime}) \right\} \right] \,
     \quad(n,m\in\wt\Lambda_R). \nn
\eea
``$\wt\Lambda_R$'' represents three dimensional half-integer lattice
which can be interpreted as a normalized root lattice spanned by the
normalized even roots $\{\wt\alpha_{(i)}\}$. 
According to this interpretation, 
a set of points on the lattice is thought to be
$\{\sum_{\wt\alpha\in\wt\Delta^+} n_\alpha \wt\alpha | n_\alpha\in\Z \
for\ \wt\alpha\in\wt\Delta_0^+ ; 
n_\alpha = \{-1/2,0,1/2\} \ for\ \wt\alpha\in\wt\Delta_1^+\}$, 
while $\wt\Delta^+$ is a set of normalized positive roots. 
$\wt\Delta^+_0$ and $\wt\Delta^+_1$ are, respectively, the normalized
even roots and the normalized odd roots which is normalized after
projected onto the root space of the even subalgebra.
As is seen in eq.(\ref{eq:gaugeparam}), since coefficient of the
$h_\theta$ has 
to vanish in the power of exponential, we may put $n_{\alpha_\theta}$ and
$\theta$-components of odd roots be zero. 
With the expression $n+m\tau = \sum_{i=+,-,\theta}( n^i+m^i\tau )
\wt\alpha_{(i)}$, it turns into $ n+m\tau = \sum_{i=+,-}
\{(n^i-\frac12 n_1^i) + \tau(m^i-\frac12 m_1^i) \} \wt\alpha_{(i)} $, where
$(n_1^+,n_1^-),(m_1^+,m_1^-)=\{(-2,0),(0,-2),(\pm1,\pm1),(0,0),
(\pm1,\mp1),(2,0),(0,2)\}$.
It gives the next variation for the field $\Psi_k(u_a)$, taking
eq.(\ref{eq:Phi}) into account.
\bea
 \label{eq:gauge-Phi}
     \Psi_k(u_a \hat{g}_{n,0}) &=& \Psi_k(u_{a+n}) \nn\\
      &=& \exp \Bigl[ -\frac{\pi}{2 \im \tau} \sum_{i=+,-}
     \left\{ k^i \left((n^i - \frac12 n_1^i)^2 + 2 a^i( n^i - 
      \frac12 n_1^i) \right) \right\} \Bigr] \Psi_k(u_a) ,
     \nn\\
     \Psi_k(u_a \hat{g}_{0,m}) &=& \Psi_k(u_{a+m\tau}) \nn\\
      &=& \exp \Bigl[ -\frac{\pi}{2 \im \tau} \sum_{i=+,-}
     \left\{ k^i \left( \tau\bar\tau
     (m^i - \frac12 m_1^i)^2 + 2 \bar\tau a^i (m^i -
     \frac12 m_1^i)\right) \right\} \Bigr] \Psi_k(u_a) , \nn
\eea
where $k^i = \frac{-2 k}{ (\alpha_{(i)})^2}$, that is, $(k^+,k^-)=( k
\gamma^{-1} , k (1-\gamma)^{-1} )$. Note that $k= k^+ k^-/(k^+ + k^-)$.

For simplicity we restrict ourselves to special cases
$n_1^i,m_1^i=0,\pm2$, where  we obtain 
one of the solutions of above two equations 
\beq
 \label{eq:wavePsi}
    \Psi_{k,p}(\tau,a) = \exp\left\{- \sum_{i=\pm} \frac{\pi k^i }{2
        \im\tau} (a^i)^2 \right\}
    \Theta_{k^+,p_+}(\tau,a^+) \,\,\Theta_{k^-,p_-}(\tau,a^-) , 
\eeq
where $\Theta_{k,p}(\tau,a)$ are the $su(2)$-theta functions of level $k$ and
$p\in\Lambda_W / k \Lambda_R$
\bea
\label{def:su2theta}
   \Theta_{k,p}(\tau,a) \equiv \sum_{\mu\in\Lambda_R} \exp\left[\,{\rm
       i}\pi \tau k \left( \mu + \frac{p}{k} \right)^2 + 2\pi{\rm i}\,k
     \left( \mu + \frac{p}{k} \right) a\, \right] ,\nn
\eea
$\Lambda_W$ and $\Lambda_R$ being the
weight and root lattice of $su(2)$, respectively.
We can show these solutions satisfy the Gauss law constraint.
Noting that the most general solution is any linear
combination of the functions in eq.(\ref{eq:wavePsi}),
we can take a Weyl anti-symmetrized combination of the theta functions.
Together with an appropriate $\Xi(u_a)$, 
it will provide two orthogonal $su(2)$ parts of 
large $N=4$ characters revealed by Petersen et al.\cite{PT}.

\section{Conclusions and discussions}
\label{sec:4}

In this paper, we formulate the anti-holomorphic quantization of
$D(2|1;\alpha)$ CSGT, giving an inner product of the Hilbert states
and the Jacobian coming from a parameterization of the
gauge field. 
After that, we extend the correspondence between CSGT Hilbert states
and SCFT characters to the $n_1,m_1=0,\pm2$ cases of large $N=4$ SCFT,
using the HR constraints in
eq.(\ref{def:HR}). 
If we clarify a full contribution of the odd parts, $n^i_1$, $m^i_1$ 
in $\wt\Lambda_R$, a 
requirement of the modular invariance on the inner product will
establish an entire identification of the characters and the Hilbert
states. It will lead to form a modular
invariant combination of large $N=4$ characters at the same time.
With those of large $N=4$ characters, Wilson line operators lying on $\Sigma$ construct braiding and fusion
matrices on them and provide link invariants on $\mani$. Finally
we remark that this
procedure will also be applicable to the case of small $N=4$ SCFT. 
These will be discussed elsewhere\cite{ishimoto}.

\vskip8mm

\ni
{\Large {\bf Acknowledgments}}

\medskip

The author is grateful to S.Matsuda and T.Uematsu for
their comments and suggestions. I am grateful to K.Sugiyama for useful 
discussions and reading my manuscript in the final version.
I also thank K.Hirata and M.Kazuhara for their hospitality during my
stay in Tokyo where part of this work was done.


\end{document}